\documentclass[10pt,a4paper,final]{article}
\usepackage[utf8]{inputenc}
\usepackage{amsmath}
\usepackage{amsfonts}
\usepackage{amssymb}
\usepackage{graphics}

\title{The Tap code - a code similar to Morse code for communication by tapping}

\author{
        Stephan Raf\/ler \\
        Nürnberg, Germany \\
        frldmr@web.de
}

\date{\today}

\begin{document}

\maketitle

\begin{abstract}

A code is presented for fast, easy and efficient communication over channels that allow only two signal types: a single sound (e.g.\ a knock), or no sound (i.e.\ silence).
This is a true binary code while Morse code is a ternary code and does not work in such situations.
Thus the presented code is more universal than Morse and can be used in much more situations.
Additionally it is very tolerant to variations in signal strength or duration.
The paper contains various ways in which the code can be derived, that all lead to the same code.
It also contains a comparison to other, similar codes, including the Morse code, in regards to efficiency and other attributes. The replacement of Morse code with Tap code is not proposed.

\end{abstract}

\section{Introduction - the Polybius square and the Morse code}\label{secintro}

The communication by means of codes is universal and ubiquitous, it is human condition and will never go away, even with the modern devices we now have. The ancient greeks devised one such communication scheme for the transmission of messages over channels that are not reachable by natural means like human speech or writing: the Polybius square \cite{polybius}. The greek alphabet has 24 letters and they are fitted in a 5 by 5 scheme. Every letter is identified by two numbers from 1 to 5, the coordinates of the letter in this scheme. These two numbers can now be transmitted e.g.\ by waving flags or lighting fires, which is visible over a long distance. The two numbers can also be knocked - a short pause is needed between them of course - and this has been used e.g.\ in prisons in war by POWs to keep up morale \cite{tapcode}. Usage of this code can be tedious however: it is not optimized, i.e.\ not adapted to letter frequencies.

Here the Morse code enters the scene \cite{morsecode}. The fact that more frequent letters should get shorter codes than less frequent ones, has been known for a long time, and Morse used this to devise a very efficient code to send messages over telegraph lines. He never intended for this code to be used with knocking however. It is the impression of the author that this fact has been a misconception for a long time, and that this needs now to be clarified and fixed once and for all. The Morse code is not the universal, generic, binary code. In fact it is not even binary, but ternary. This may be surprising for some readers, so we will explain this now: the fact that Morse works with an "on" and an "off" state does not mean, that it is binary. Because with this information alone, it is not clear, for how long "on" should be meaningful. The time dimension is the important one here. Many codes could be defined by letting different "on" times have different meanings. Morse chose to have two: a short one (dit) and a long one (dah). And then there is a third one: the pause or silence. These are three different signals with three different meanings. Morse could have as well defined a code with only two different signals: short (dit) and pause. This is what a knock or tap code looks like. The short signal can be anything: a dit, a knock, a tap, a touch, a blink of light, etc.

This leads us now to the creation of a tap code from scratch, that will combine the advantages of both described codes, the Polybius square and the Morse code.

\section{Tap code}\label{sectapcode}

Just like Morse we will rely on the fact that every code word has to start with a signal (not with a pause). We will denote this as "1" in the following, and a pause as a "0". The first code that comes to mind is depicted in (Table \ref{tab:tabeinfach}).

\begin{table}[h]
\caption{a very simple tap code}
\begin{center}
\begin{tabular}{ccl}
  e & = & 10 \\
  n & = & 110 \\
  i & = & 1110 \\
  r & = & 11110 \\
  s & = & 111110 \\
  t & = & 1111110 \\
  a & = & ... \\
\end{tabular}
\end{center}
\label{tab:tabeinfach}
\end{table}

Here the frequencies of the letters in the german language has been used to assign longer codes to letters the less frequent they are. This code has the disadvantage (for a practical tap code) that the counting of the number of knocks becomes more and more tedious if not impossible for less frequent letters.

Thus the single pause "0" can not be used, and the next longer pause is "00", which is double the length. We will have of course a basic time interval just like in Morse code. Basic time intervals are universal and ubiquitous, see for instance music. We will consider the musical analog for a moment: a tap, or "1" would correspond to a note, of a certain time value, e.g.\ a so called sixteenth. A pause or "0" would correspond to a rest, which is described by the same time values. So in musical language, "00" would correspond to an eighth rest. If we think about this for a moment, it becomes clear that it would be very sensible to let our code be of such a form, that it is adaptable to an eighth beat (as opposed to a sixteenth beat). This statement is of course not scientifically, rigorously justifiable. But an objective conclusion can nonetheless be reached just by considering the following: every musician will agree, what is easier to play and to hear. At least this is the conviction of the author (who is only an amateur musician). For the design of our code we have now an easy option and a hard(er) option, and we will choose the easy one.

So all bit sequences will be of even length in order to fit into the eighth beat form. Efficiency concerns one might have at this point are considered less important than the eighth beat.

The double pause allows us now to have much more flexibility between the start of the code word "1" and the end "00". Of course we can not use "00" anywhere, since this would mark the end. The last bit of this in between sequence will have to be "1", since if it were a "0", this would just mean that we have moved the end one bit to the left, i.e.\ shortened the code word by one bit. So all in all we will look now at a code that is constructed by the following rules:

\begin{enumerate}
\item the bit sequence that is assigned to a letter starts with "1", ends with "1", and has "00" appended as rest and end marker
\item if the bit sequence is of odd length, an additional "0" is appended to let the following code word start  at an eighth beat position again
\end{enumerate}

The code that is constructed with these rules, the proposed Tap code, is this (Table \ref{tab:tapcode}):

\begin{table}[h]
\caption{the Tap code}
\begin{center}
\begin{tabular}{cclccl}
  e & = & 10      & f & = & 111011 \\
  n & = & 11      & w & = & 110111 \\
  i & = & 1010    & k & = & 101111 \\
  r & = & 1110    & z & = & 111111 \\
  s & = & 1101    & p & = & 10101010 \\
  t & = & 1011    & v & = & 10110110 \\
  a & = & 1111    & ä & = & 11101010 \\
  h & = & 101010  & ü & = & 10111010 \\
  d & = & 111010  & ß & = & 10101110 \\
  l & = & 110110  & ö & = & 10101011 \\
  u & = & 101110  & j & = & 11011010 \\
  c & = & 110101  & x & = & 10101101 \\
  m & = & 101101  & y & = & 10110101 \\
  g & = & 101011  & q & = & 11010110 \\
  o & = & 111110  & . & = & 10111110 \\
  b & = & 111101  & ? & = & 11101011 \\
\end{tabular}
\end{center}
\label{tab:tapcode}
\end{table}

The end marker "00" has been omitted for clarity. We see that some codes end with "0" and some with "1", as for the ones with "0", rule 2 has been applied. The question is now, why exactly these codes, and are there any others that were left out, and if yes, why. In order to clarify this, we will proceed very systematically simply by considering all possible bit sequences of a certain length and filtering those out, that we cannot use for the code, i.e.\ that have "00" or longer sequences of "0" in them. The remaining sequences will have a number of "1"s in them and when we have the opportunity, we will choose those with a minimum number of "1"s. This is due to the fact that sending a "1" might be an energy consuming operation in some situations, and we want to minimize this energy.

\section{Construction}\label{seccnst}

Here is now the most basic, brute force but systematic way of the construction:
\\

length 1: 1

This becomes "10", the code for "e", since a "0" is appended. 
\\

length 2: 11

This is our "n".
\\

length 3: 101, 111

becomes

1010  i

1110  r
\\

length 4:

1001

1011  t

1101  s

1111  a

The first code word is dropped, since it contains "00".
We realize, that we need only to consider the inner bits, since the outer ones are always "1".
For length 5 (i.e.\ 6 with appended "0") there are 3 inner bits, and thus 8 possible codes:
\\

length 5: 000, 001, 010, 011, 100, 101, 110, 111

Three of them are dropped, from the remaining we get

101010  h

101110  u

110110  l

111010  d

111110  o

Well, but what about g, m and c? They are coming next.
These codes here are with an eighth at the end. The following ones all have a sixteenth at the end.
\\

length 6:
0000,
0001,
0010,
0011,
0100,
0101 *,
0110 *,
0111 *,
1000,
1001,
1010 *,
1011 *,
1100,
1101 *,
1110 *,
1111 *

Codes that are taken into account are marked with *. These give

101011  g

101101  m

101111  k

110101  c

110111  w

111011  f

111101  b

111111  z

So far we have left none out and we have obtained all code words of length smaller than four eighths.
The codes with 5 inner bits are now 32 by number and we will not need all of them any more.
We will take those with a minimum of "1"s in them.
\\

length 7:
00000,
00001,
00010,
00011,
00100,
00101,
00110,
00111,
01000,
01001,
01010 *,
01011 *,
01100,
01101 *,
01110 *,
01111 *,
10000,
10001,
10010,
10011,
10100,
10101 *,
10110 *,
10111 *,
11000,
11001,
11010 *,
11011 *,
11100,
11101 *,
11110 *,
11111 *

Due to the "00" rule the number of actually remaining words is quite small.
We get (again with an eighth at the end):

10101010  4 p

10101110  5 ß

10110110  5 v

10111010  5 ü

10111110  6 .

11010110  5 q

11011010  5 j

11011110  6

11101010  5 ä

11101110  6

11110110  6

11111010  6

11111110  7

The number of "1" bits is included now, in order to confirm that we have indeed chosen the combinations with the lowest number. Now we still need some of the combinations with length 8 that give the codes with sixteenth at the end.
\\

length 8:
000000,
000001,
000010,
000011,
000100,
000101,
000110,
000111,
001000,
001001,
001010,
001011,
001100,
001101,
001110,
001111,
010000,
010001,
010010,
010011,
010100,
010101 *,
010110 *,
010111 *,
011000,
011001,
011010 *,
011011 *,
011100,
011101 *,
011110 *,
011111 *,
100000,
100001,
100010,
100011,
100100,
100101,
100110,
100111,
101000,
101001,
101010 *,
101011 *,
101100,
101101 *,
101110 *,
101111 *,
110000,
110001,
110010,
110011,
110100,
110101 *,
110110 *,
110111 *,
111000,
111001,
111010 *,
111011 *,
111100,
111101 *,
111110 *,
111111 *

This gives

10101011 5 ö  

10101101 5 x  

10101111 6    

10110101 5 y  

10110111 6    

10111011 6    

10111101 6    

10111111 7    

11010101 5 (here we observe an unused code with 5 "1"s. for why, see below)

11010111 6    

11011011 6    

11011101 6    

11011111 7    

11101011 6 ?  

11101101 6    

11101111 7    

11110101 6    

11110111 7    

11111011 7    

11111101 7    

11111111 8    

All the letters already have codes with 5 "1"s. So for "." and "?" it was the opinion of the author, that two codes with 6 "1"s could be used. The remaining unused 5 "1" code is a very long syncopation in musical terms, and has been avoided intentionally to not unnecessarily make the keeping of the rhythm harder.
\\

Now that we have seen the explanations and illuminated the reasons for all the design choices that have been made, other ways of construction that lead to the same end result might be of interest. This may shed more light on why the code is like it is, and why it can be viewed as being universal in a sense.
\\

Deriving the Tap code from the Morse code:
\\

Knock all Morse code words, i.e.\ all possible combinations of "dit" and "dah", "$\cdot$" and "$-$", short and long, sixteenth and eighth. If two are sounding the same, take the one of them that fits into the eighth raster.
Discard the other one. Examples are "$- - \cdot$" and "$- - -$" which sound the same when knocked, but only "$- - -$" fits into the eighth raster, so "$- - \cdot$" is dropped. While "$- \cdot \cdot$" and "$- \cdot -$" also sound the same, but now "$- \cdot \cdot$" is taken. This way, at the end of the code word, sometimes a sixteenth is converted to an eighth and sometimes an eighth is converted to a sixteenth. So one could say, the Tap code \textit{is} the Morse code, only without the ambiguity at the end.
\\

Deriving the Tap code from the Polybius square code:
\\

In the introduction the Polybius square code was described as being able to be tapped, simply by tapping the two coordinates of a letter in the square. When one does this, it actually already sounds quite like the Tap code. The difference lies in the fact that in the Polybius square code there is always one pause to separate the coordinates, they are all of the form x,y (with , denoting the pause). If we view the Tap code in this scheme, we observe different numbers of pauses, and even some words with \textit{no} pause.
\\

If we use these observations for a construction method, we would start with exactly these code words that have no pause: let them be denoted simply 1 \; 2 \; 3 \; 4 \; 5 \; ...
\\

It is clear how to proceed: now come all the combinations with two numbers:
 1,1 \; 1,2 \; 2,1 \; 1,3 \;  2,2  \; 3,1 \; ... 
\\

And then all the ones with three numbers, four, and so on. What remains to do then is to order them by overall length in bit notation. This will give exactly the set of bit sequences that is needed for the Tap code. The Tap code in this notation looks like this (Table \ref{tab:tapcodepoly}):

\begin{table}[h]
\caption{the Tap code in Polybius square notation}
\begin{center}
\begin{tabular}{cclccl}
  e & = & 1       & f & = & 3,2      \\
  n & = & 2       & w & = & 2,3      \\
  i & = & 1,1     & k & = & 1,4      \\
  r & = & 3       & z & = & 6        \\
  s & = & 2,1     & p & = & 1,1,1,1  \\
  t & = & 1,2     & v & = & 1,2,2    \\
  a & = & 4       & ä & = & 3,1,1    \\
  h & = & 1,1,1   & ü & = & 1,3,1    \\
  d & = & 3,1     & ß & = & 1,1,3    \\
  l & = & 2,2     & ö & = & 1,1,1,2  \\
  u & = & 1,3     & j & = & 2,2,1    \\
  c & = & 2,1,1   & x & = & 1,1,2,1  \\
  m & = & 1,2,1   & y & = & 1,2,1,1  \\
  g & = & 1,1,2   & q & = & 2,1,2    \\
  o & = & 5       & . & = & 1,5      \\
  b & = & 4,1     & ? & = & 3,1,2    \\
\end{tabular}
\end{center}
\label{tab:tapcodepoly}
\end{table}

Deriving the Tap code from a 4 symbol alphabet representation:
\\

We have seen that the two bit pause "00" suggests a raster that lets code words be an even number of bits in length. Now two bits represent a number from 0 to 3 or a four symbol alphabet out of which one could construct another Tap code. Some rules will have to be obeyed nonetheless. The first position for a code word can only be occupied by "10" or "11", not by "01" (and of course not by "00" since this is still our pause sequence). The last position (before the pause), can however now be occupied by all three possibilities. In between we have to be careful not to use combinations like "10" "01" that would give a "00". Now this last rule could be disputed. For example for a musician, such a combination is not uncommon and not particularly hard to tap. It can be argued that some of the combinations of this form could indeed be practical to use. But the designer of the Tap code has decided to discard these for the sake of a strict "no 00 in between" rule.
\\

This concludes the presentation of different methods to construct the Tap code. Many more are conceivable and every single one of them sheds some new light on properties and alternatives of the Tap code. At some point in time or the other they have all been considered by the author. None of them have convinced him that there is something to alter at the Tap code and it has remained stable and clear as a crystal for a long time now.

\section{Analysis and comparison}\label{secanal}

For the comparison of the efficiency of different codes and to determine where the Tap code falls into this list, a short computer program has been written and used to compute empirically for a test text the average bits per character or what will in the following also be called "bit efficiency". (Really it is the reciprocal of the bit efficiency which would be "characters per bit", but this seems too abstract a notion to the author.)
\\

Letters are converted to lower case, space is considered a character, i.e.\ it gets a code and a frequency. Four german umlauts are included, everything else is ignored, so that there are 31 different code words in the end. First, a Huffman encoding is calculated, and it turns out that the particular test text can be encoded with an average of 4.2 bits per character. The Huffman code is of course not practical for humans to use for tapping.
\\

The next code is simply a 5 bit per character encoding, which is also impractical to use, because sequences of "0"s that are longer than 2 would require an extra "metronome" to keep track of the exact number. 5 bit because $ 2^5=32 $ and we have 31 characters to encode, the word 00000 is not used.
\\

Next is the Tap code which turns out to have approximately 6 bit efficiency. Considering that most code words have 3 bits fixed, i.e.\ the first bit is 1 and the last two 0, this is an intriguing result. This just shows, that most of the information of such codes flows into the separation of the single letters. Once the means for this are there, the remaining encoding can be very efficient.
\\

With Morse code there is a slight problem: the pure bit efficiency is very poor, 8.26 bit, but: in practice, a dit would correspond to a tap in the Tap code, so for comparing efficiency in reality, one would have to halve the bit efficiency. This is due to the fact that dit and dah are really two different sounds, and the premise of the Tap code was that we can not have that, but only one single sound. So, the author is not quite sure how to interpret this result, maybe this just shows that Morse and Tap code are not really comparable. However, one can do the following calculation: $ 2^x=3^{4.13}$  which leads to 6.5 bits. This would make the Tap code even more efficient than the Morse code. But only in an abstract information theoretical sense. In practice the 4.13 ternary efficiency holds, and this is much more efficient than the Tap code. To emphasize it once more: this is no argument against the Tap code, because the premises are completely different.
\\

The Polybius square in its original form has an efficiency of around 7.7 bit and this compares well with the Tap code. It can be made more efficient by assigning the short codes to the more frequent letters, and by using the diagonals which correspond to constant code length instead of just filling the square. Then the efficiency increases to 6.5 bit which is quite good, compared to the 6 bit of the Tap code. Note however that this optimized Polybius square tap code is not very much easier to learn than the Tap code proposed in this paper. So if one is willing to learn a tap code, the proposed Tap code is still the best choice. Furthermore it has the advantage over the Polybius one, that it is completely based on even numbers of taps. The original Polybius square does not accommodate all 26 letters of the latin alphabet. The optimized version does and has still room for more. But in both variations an extension leads to long sequences of taps which gets tedious soon. In the Tap code the longest sequence is 6 for the letter z, which is the maximum that should ever be used in the opinion of the author.
\\

The investigation has also been done with an english text, and since the Tap code is adapted to german letter frequencies and the Morse code to english letter frequencies, there is a slight difference up or down on the order of 0.1 bit respectively. But this is not critical for the overall performance comparison of the codes. The relative efficiencies stay roughly the same, and the Tap code could easily be adapted to an international version of letter frequencies.

\section{Discussion}\label{secdisc}

In this section we will reply in advance to some arguments that have been made in the past communications of the author with various other people (and I want to thank all these people for these discussions, and for trying the Tap code with me in practice, especially my wife and my elder daughter). This will be a lose collection.
\\

It has been noted in private communications, that the Tap code is similar to the Fibonacci code. This is true, but it is a corollary, and was not the starting point of the construction of the Tap code. In fact, the same statement could be made about the Morse code, and probably many other things. This is not because everyone searches for applications of the Fibonacci code (because indeed, very few people even know about it), or the Fibonacci numbers. But this is because the Fibonacci numbers are ubiquitous and appear everywhere naturally, anyway. To turn this fact around and say the Fibonacci code has been the starting point for it all, is not of scientific integrity, it is not true. A musician who writes a piece of music with consecutive eighth and sixteenth notes has not done so because of the Fibonacci code, so would you tell him, that he has found an interesting application of the Fibonacci code? How far would you go with this? Would you tell Einstein that with his theory of General Relativity he has found an interesting application of addition and multiplication? Would you tell God or the Evolution that with the nautilus shell they have found an interesting application of the Fibonacci code? No, the Fibonacci numbers come after the fact. As a physicist, the author has a more down to earth world view.
\\

When the condition of even code word lengths is not used, a code is obtained that can be used for data encoding on a computer. The most frequent letter gets in this case the code "0" (!). All other codes start with "1", and the second most frequent letter would get the code "100". Then "1100" and so on using the construction rule 1. This code (with "0" being space between words) is harder to tap and is thus not proposed as the generic Tap code.
\\

The Tap code is not harder to learn than the Morse code, if anything, it is easier.
\\

The Tap code can not be tapped as fast as Morse since successive short signals appear to the human ear more as a frequency than single, distinguishable sounds, below a certain time interval. In Morse code this is not so much of a problem, because the number of consecutive short signals, i.e.\ dit's is lower, and is interleaved with longer dah's, which is more pleasing and gives a richer structure to the human ear. In this regard (speed) the Tap code is inferior to the Morse code, it has other strengths.
\\

For an international version of the Tap code the letter frequencies would be different, but the set of bit sequences would be the same. The german umlaut section could be replaced with national special characters, or frequent diphthongs like "th" in the english language. The digits could be represented by e,n,r,a,o,z obviously for the numbers 1-6, and then s for 7, t for 8, h for 9 and i for 0. It would be clear from the context that not a letter is meant, but a number.
\\

The author is under the impression that the Morse code is perceived as "natural" while the Tap code appears "unnatural", "artificial", "constructed" or similar attributes in this vein. This is not true. The Morse code is not more or less constructed than the Tap code. In fact, to the author, if anything the Tap code appears more natural, since it can be practiced without any technical devices, just with a finger on a wooden table or indeed by clapping hands. If you can demonstrate communication by clapping hands with the Morse code you might be justified to say that it is equally natural than the Tap code. Until then this is doubted. Never, however, will the Morse code be \textit{more} natural, general, universal or robust than the Tap code.
\\

It is not necessary to adhere strictly to the exact beat. Indeed it is sufficient to recognize the numbers of consecutive taps to identify letters. So for instance "m" would be identified by a sequence of 1, then 2, and then again 1 knocks. Other examples are "l" as 2,2 or "d" as 3,1 or "e" as 1 and so on. For this, the exact rhythm is not required, only the pause after the letter has to be longer in this case to identify unmistakeably the end of the code word. With perfect rhythm the minimal pause is "00" and then the Tap code is very fast and efficient.
\\

It makes no sense in the opinion of the author to use the Tap code (or the Morse code) in a written form, because all these codes are meant to be used in a time resolved, rhythm based manner. So while writing little "secret" messages with dots on post cards might be a fun application, this is not what the code was designed for.
\\

The overall speed of the Tap code is determined by the time interval, i.e.\ how long a sixteenth note takes  measured e.g.\ in fractions of a second. As we have seen that the average bit efficiency is about 6 bit, a time interval of 1/6th of a second would lead to an average speed of 1 letter per second or roughly 10 words per minute which is compared to the speed that most people reach with Morse code not bad.
\\

How does one discern the time interval, i.e. how are i and n, r and h, a and p distinguished, or perhaps even those from the respective number of e's? Well, this is only a problem if the receiver has never heard an other letter from the source before. As soon as the receiver hears a letter like s or t, where both lengths occur, everything is clear. This is no different in Morse code, if the duration of dit's and dah's is not kept very precise. Then a slow s could easily be mistaken for a very fast o. But only if one can make absolutely no assumptions about the speed performance capabilities of the source, and in practice, in reality this is almost never the case.

\section{Conclusion}\label{secconc}

We have presented the Tap code, a code that is similar to Morse code, but can be used in more situations since it is a true binary code that relies only on a single signal and a pause. We have explored the efficiency and it has turned out to be more efficient in a bit based information theoretical sense than Morse and other codes. But since this does not apply to the practical use situations of the Morse code, the Tap code will not replace it, and was never intended to. The Morse code is perfectly adapted to its channel, the telegraph line. The Tap code always was intended to be useful in situations where Morse code can not be used due to physical channel restrictions, i.e.\ when one can only knock or tap or give a single signal in any form. It is the hope of the author that it will find many useful and fun (and maybe even sometimes desperate) applications that he cannot possibly imagine at the time of the writing. What he can imagine very well, though, is the human condition, and the Tap code is for humans.


\begin{thebibliography}{3}

\bibitem{polybius} "The Histories of Polybius", published in Vol. IV of the Loeb Classical Library edition, 1922 thru 1927, public domain. 

\bibitem{tapcode}  Brace, Ernest C. (1988). "A Code to Keep: The true story of America's longest held civilian prisoner of war in Vietnam." St. Martin's Press. ISBN 0-7090-3560-8. pp. 171–172, 187–188.

\bibitem{morsecode} "International Morse code Recommendation ITU-R M.1677-1". itu.int. International Telecommunication Union. October 2009.

\end{thebibliography}
\end{document}